\documentclass[cits]{PoS}
\usepackage{epstopdf}

\title{The mystery of the missing GRB redshifts}

\ShortTitle{The mystery of the missing GRB redshifts}

\author{\speaker{David M. Coward}$^{1,2}$, 
Eric Howell$^{1}$,
Marica Branchesi$^{3,4}$, Dafne Guetta$^{5,6}$ and Chadia Kanaan$^{7}$\\
 $^{1}$School of Physics, University of Western Australia, Crawley WA 6009, Australia\\
$^{2}$Australian Research Council Future Fellow\\
$^{3}$DiSBeF - Universit\`a degli Studi di Urbino `Carlo Bo', I-61029 Urbino, Italy\\
$^{4}$INFN, Sezione di Firenze, I-50019 Sesto Fiorentino, Italy\\
$^{5}$Department of Physics and Optical Engineering, ORT Braude, P.O. Box 78, Karmiel, Israel\\
$^{6}$ INAF - Osservatorio Astronomico di Roma, Via Frascati 33, I-00040 Monteporzio Catone (Roma), Italy\\
$^{7}$University of Nice-Sophia Antipolis \& Observatoire de la C\^ote d'Azur\\
   Laboratoire Lagrange, UMR 7293, BP 4229, F-06304, Nice Cedex 4, France\\
 E-mail: \email{david.coward@uwa.edu.au}, \email{eric.howell@uwa.edu.au}, \email{marica.branchesi@uniurb.it}, \email{dafne.guetta@oa-roma.inaf.it}, \email{Chadia.Kanaan@oca.eu}}

\abstract{It is clear that optical selection effects have distorted the ``true" GRB redshift distribution to its presently observed biased distribution. We constrain a statistically optimal model that implies GRB host galaxy dust extinction could account for up to 40\% of missing optical afterglows and redshifts in $z = 0-3$, but the bias is negligible at very high-$z$. The limiting sensitivity of the telescopes, and the time to acquire spectroscopic/photometric redshifts, are significant sources of bias for the very high-$z$ sample. We caution on constraining star formation rate and luminosity evolution using the GRB redshift distribution without accounting for these selection effects. }

\FullConference{Gamma-Ray Bursts 2012 Conference -GRB2012,\\
		May 07-11, 2012\\
		Munich, Germany}

\begin{document}

\section{GRB optical selection effects}
We define gamma ray burst (GRB) optical afterglow (OA) selection effects as the combination of sensitivity limited optical follow-up and phenomena, astrophysical and instrumental, that reduce the detection probability of an OA. Some of the more widely understood effects are discussed by Fynbo et al.\cite{2009ApJS..185..526F} in detail. Optical biases have reduced the fraction of {\it Swift} triggered OAs, and have introduced a selection towards detecting the brightest OAs, hence the more nearby bursts. Additionally, there are biases that distort the redshift distribution over certain redshift ranges (see e.g. \cite{cow08,cow09}):\\
\begin{enumerate}

\item {\bf Malmquist bias:} This bias arises because the telescopes and instruments acquiring OA absorption spectra (and photometry) are limited by sensitivity. In reality, the instruments acquiring redshifts are biased to sampling the bright end of the OA luminosity function. To account for this bias, it is necessary to have some knowledge of OA luminosity function (which is uncertain especially at the faint end), and an estimate of the average sensitivity limit of the instruments. This is the most fundamental bias that encompasses all flux limited detection and is the basis for modelling a selection function for OA/redshift measurement.

\item {\bf Redshift desert:} The so-called redshift desert is a region in redshift ($1.4 < z < 2.5$) where it is difficult to measure absorption and emission spectra. 
As redshift increases beyond $z\sim 1$, the main spectral features become harder to recognize as they enter a wavelength region where the sensitivity of CCDs starts to drop and sky brightness increases. Beyond $z\sim1.4$, the spectral features move beyond 1 $\mu$m, i.e., into the near-IR.  In the case of actively starforming galaxies at $z> 1.4$, these are several narrow absorption lines over the UV continuum, most of which originate in the ISM of these galaxies. 

%

\item {\bf Different redshift measurement techniques:} Historically, because of the deficiency in pre-{\it Swift} ground-based follow-up of GRBs, there was a strong bias for imaging the brightest bursts. Because the brightest bursts are predominantly nearby, a significant fraction of the first GRB redshifts were obtained by emission spectroscopy of the host galaxy.
In the {\it Swift} era (from 2005 onwards), an optical afterglow (OA) is usually required to measure a redshift. For most high-$z$ GRBs, this is achieved by absorption spectroscopy of the GRB afterglow. The host galaxies are usually too faint to make a significant contribution to the spectra. Most GRB spectroscopic redshifts are acquired by large aperture ground based telescopes, including VLT, Gemini-S-N, Keck and Lick (see \cite{2009ApJS..185..526F} for a more complete list along with specific spectroscopy instruments). The measurement of a GRB redshift depends strongly on the limiting sensitivity and spectral coverage of the spectroscopic system. This bias is expected to manifest at high-$z$, where the optically brightest OAs are near the limiting sensitivity of the telescope. 

\item {\bf Host galaxy extinction:} 
 There has been growing evidence that dark bursts are obscured in their host galaxies e.g. \cite{jakob04,2006ApJ...647..471L,2009AJ....138.1690P,2012MNRAS.tmp.2447S}. The detection of the near-infrared (IR) OAs of some GRBs (which would have been considered as dark bursts because their OAs were not detected in any bluer bands) provides evidence for dust obscuration \cite{tanvir08}. These studies generally show that GRBs originating in very red host galaxies always show some evidence of dust extinction in their afterglows. Also a significant fraction of dark burst hosts have extinction columns with $A_V \sim1$ mag, and some as high as $A_V =2-6$ mag  \cite{2009AJ....138.1690P}. Rossi et al. \cite{rossi12} performed a search for the host galaxies of 17 bursts with no optical afterglow. They find in seven cases extremely red objects in the error circles, at least four of them might be dust-enshrouded galaxies.\\

\end{enumerate}

\section{Malmquist bias for GRB redshift measurement}
A telescope limiting magnitude ($m_\mathrm{L}$) defines a threshold for obtaining a redshift. Because of the different instruments engaged in spectroscopic redshift measurement it is difficult to constrain this important parameter. For definiteness, we take $m_\mathrm{L} = 24$, approximating an average limit for the telescopes used to acquire the redshifts for the faintest bursts.

With a reference time $t_\mathrm{c}$ of 1 day, the OA limiting luminosity can be calculated at time $T_\mathrm{z}$ \cite{imerit09}, assuming all OAs fade in luminosity by $\alpha\propto t^{-1}$: 
\begin{equation}
\label{eqn:Pogson}
M_\mathrm{L}(z) = m_\mathrm{L} - 5\log_{10}\Big(\frac{d_\mathrm{L}}{10}\Big) - \frac{5\alpha}{2} \log_{10}\Big(\frac{T_\mathrm{z}}{t_\mathrm{c}}\Big) + \delta m(z).
\end{equation}
We use $T_\mathrm{z}\approx400$ min, which is the mean time taken to acquire a redshift \cite{cow09}, and $\delta m(z)$ accounts for a possible increase in luminosity with redshift from either intrinsically brighter sources or less dust obscuration in the hosts. 
Using the above definition, and OA luminosity function $\varphi(M)$, the OA selection function defines the fraction of observable OAs with redshift:
\begin{equation}
\label{select2}
\psi_{\rm OA}(z) = \int_{M_\mathrm{L}(z)}^{M_\mathrm{Max}}{\varphi(M)} \mathrm{dz}\;.
\end{equation}
\section{Results}
Equation \ref{select2} can be linearly combined with other completeness functions for dust extinction and the redshift desert. For the latter, we employ simple scaling functions that reduce the completeness over certain redshift ranges. The GRB redshift probability distribution function, that includes the above selection effects, can be expressed as: 
\begin{equation}\label{pdf1}
P(z) = C\frac{dV(z)}{dz}\frac{e(z)}{(1+z)} \psi_{\rm Swift}(z)\psi_{\rm OA}(z)\psi_{\rm Desert}(z)\psi_{\rm Dust}(z)
\end{equation}
where $C$ is a normalization constant, $dV(z)/dz$ is the volume element for a flat-Lambda cosmology and $e(z)$ is the dimensionless source rate density evolution function (scaled so that $e(0)=1$). We assume that $e(z)$ tracks the star formation rate history.
\begin{figure}
\centering
\includegraphics[scale=0.8]{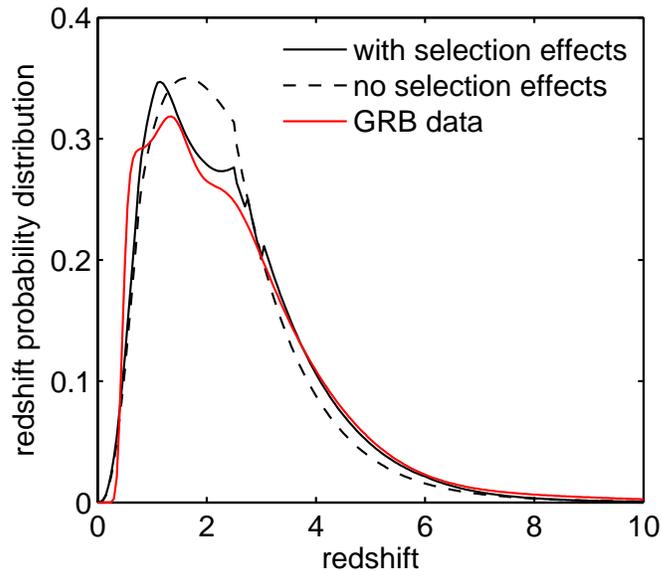}
\caption{The curves plot the expected redshift distribution for models with and without optical selection effects, and the normalized distribution of 136 {\it Swift} triggered absorption spectroscopic redshifts from the OA. The model curve (with selection effects) is constrained by a KS probability $>95$\%. } \label{fig_pdf_results}
\end{figure}

\section{Summary}
Figure \ref{fig_pdf_results} plots the expected redshift distribution including these completeness functions.
The optimal model based on a Kolmogorov Smirnov (KS) probability $>95\%$ is one that includes selection effects. {\bf Dust extinction} produces a $(30-40)$\% reduction in redshifts in $z=0-3$. The fraction of missed redshifts from the {\bf reshift desert} is $<25$\% in $z=1.5-2.5$. At high-$z$ the {\bf OA Malmquist bias} results in 20\% of redshifts missed in $z=0-5$ and increases to 50\% missed out to $z=10$.

\end{document}